\documentclass[twocolumn,superscriptaddress,aps,floatfix]{revtex4-1}

\usepackage{graphicx}
\usepackage{float}
\usepackage{bm}
\usepackage{dcolumn}
\usepackage{multirow}
\usepackage{upgreek}
\usepackage{amsmath}
\usepackage{amsfonts}
\usepackage{amssymb}
\usepackage{hyperref}
\hypersetup{colorlinks=true, linkcolor=blue, citecolor=blue, urlcolor=blue, unicode=false}

\begin{document}

\title{Free-standing magnetic nano-membranes for electron spin-filtering applications}

\author{T. \"{O}vergaard}
\affiliation{KTH Royal Institute of Technology, Materials Physics, Isafjordsgatan 22, SE-164 40 Kista, Sweden}

\author{B. M. Wojek}
\affiliation{KTH Royal Institute of Technology, Materials Physics, Isafjordsgatan 22, SE-164 40 Kista, Sweden}

\author{M. Leandersson}
\affiliation{Lund University, MAX IV Laboratory, P.O. BOX 118, SE-221 00 Lund, Sweden}

\author{H. Ohldag}
\affiliation{Stanford Synchrotron Radiation Laboratory, P.O. Box 20450, Stanford, California 94309, U.S.A.}

\author{S. Bonetti}
\affiliation{Department of Physics, Stockholm University, 106 91 Stockholm, Sweden}

\author{T. Wiell}
\affiliation{Scienta Omicron, Vallongatan 1, SE-752 28 Uppsala, Sweden}

\author{O. Tjernberg}
\email{oscar@kth.se}
\affiliation{KTH Royal Institute of Technology, Materials Physics, Isafjordsgatan 22, SE-164 40 Kista, Sweden}

\begin{abstract}
Free-standing ferromagnetic nano-membranes with thicknesses below 10 nm could effectively be used for spin selective filtering of electrons. Such membranes can work both as spin detectors in electron-spectroscopy, -microscopy and -diffraction as well as a source of spin polarized electrons. Theoretical studies and previous work has indicated that ferromagnetic membranes of a few nm  have Sherman functions in the 30-60 \% range and would provide an effective alternative to current spin detection technology. Here we demonstrate the fabrication of gold capped Co nano-membranes with a 2.6 nm Co layer and a total thickness below 10 nm. The membranes have a Sherman function $S\approx 0.41$ and a transmission of $3.7\times 10^{-2}$ for electron energies of 2 eV. The integration of such spin-filtering membranes in a hemispherical electron analyzer is shown to provide massively parallel detection capabilities and a "2-dimensional" figure of merit $FOM_\mathrm{2D}$ = 67.2, the highest reported to date.

\end{abstract}


\maketitle

\begin{figure*}[htbp]
\centering
\includegraphics[width=0.9\textwidth]{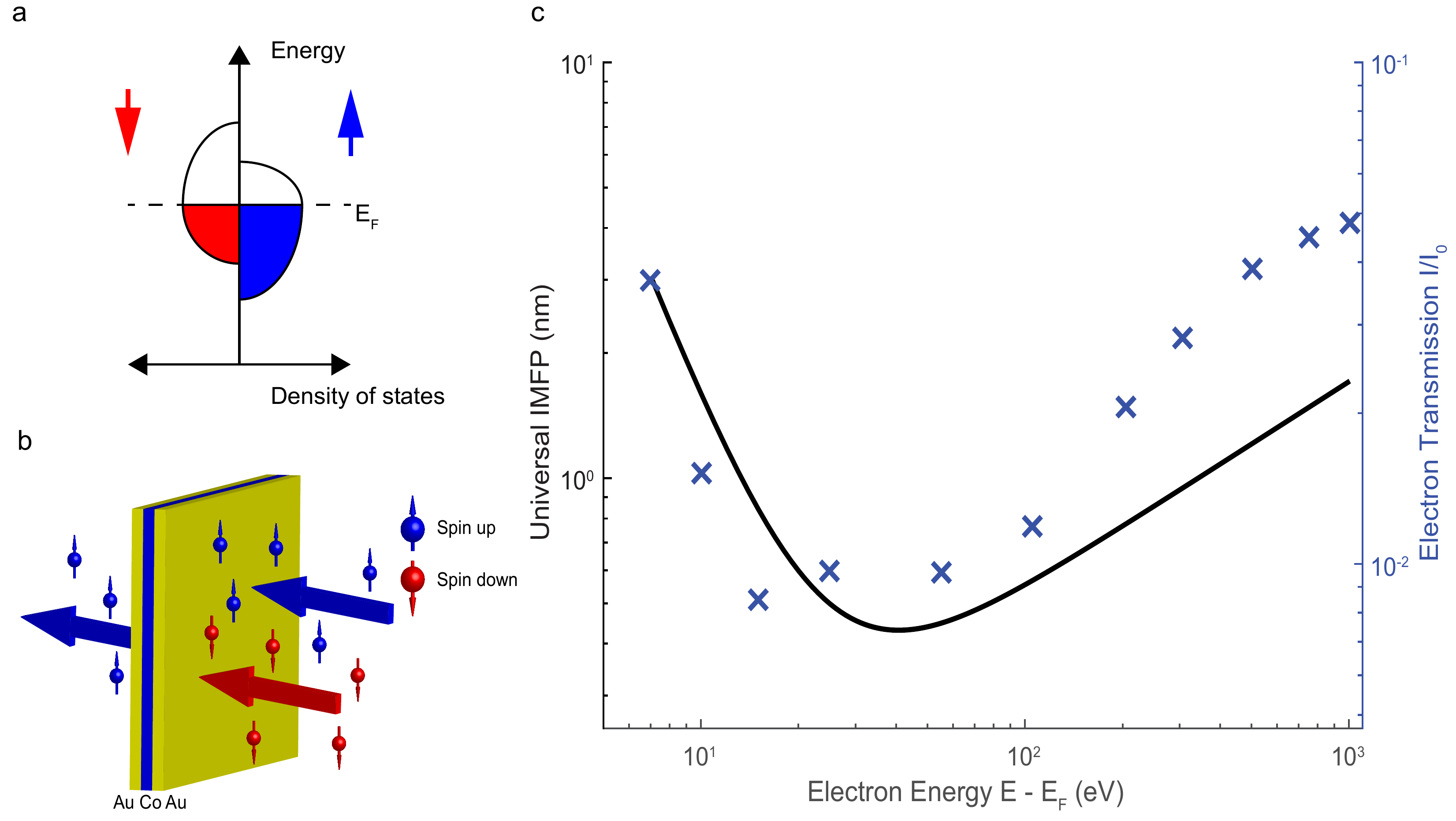}
\caption{\footnotesize Fundamentals of the spin detector. a) A schematic illustration of spin polarized valence electrons in a magnetized material. The number of available states to scatter into vary depending on the spin of an incoming electron. b) The basic design of the free-standing membrane with an in-plane magnetized Co film capped by Au to prevent oxidation. Blue represents the magnetization direction and the majority spin which can pass through more easily. c) A comparison between the "universal" Inelastic Mean Free Path (IMFP) of an electron (black line) and transmission through the present spin filter (blue crosses) as a function of kinetic energy.}
\label{fig:1}
\end{figure*}
In 1922 Otto Stern and Walther Gerlach performed their famous experiment proving the fundamental quantum mechanical property of spin using neutral silver atoms \cite{Gerlach1922nr1,Gerlach1922nr2,Gerlach1922nr3}. To understand the electronic properties of many complex condensed matter systems e.g. magnetic materials, correlated systems, and topological insulators \cite{Dziawa2012,Hsieh2008,Hsieh2009,Moore2009,Moore2010,Hoesch2004,Veenstra2014,Manchon2015}, the ability to measure spin has become crucial. Unfortunately, for the charged electron, the use of an inhomogeneous magnetic field to separate the spins does not work due to Lorentz forces acting on the electric charge \cite{Mott425}.

Since 1922, several ways of detecting electron spin have been demonstrated. One efficient technique is exchange scattering \cite{SiegmannBook} where electrons are selectively scattered by the spin polarized valence electrons in a magnetic material (Fig. \ref{fig:1}a). The scattering cross-section in a magnetic material is spin dependent due to the difference in available phase space volume of empty states for a particular spin direction. This effect can be exploited for spin detection either in reflectivity or in transmission.

The goal of any spin detector is to achieve a high figure of merit (FOM),
\begin{align}
FOM = S^2\times I/I_0 ,
\label{eq:FOM}
\end{align}
with $I$ being the transmitted intensity, $I_0$ the incoming intensity and $S$ the Sherman function \cite{KesslerBook}. The Sherman function is a measure of how well the asymmetry, $A=SP$, can be measured given a certain polarization, $P$, of the incoming current \cite{KesslerBook}. Hence, a perfect spin filter would have both a Sherman function, and a FOM equal to one.

Some of the most extensive work done on electron transmission through thin ferromagnetic films comes from H.C. Siegmann and co-workers \cite{Schönhense1993,Oberli1998,Pappas1991,Lassailly1994}. By letting a spin polarized electron current pass through a magnetized Co film, the transmission for parallel and anti-parallel spin versus magnetization could be determined (Fig. \ref{fig:1}b). Although large Sherman functions where demonstrated, the exponential decline of transmission with increasing film thickness resulted in very low transmission $I/I_0$. This prohibited the FOM from reaching useful values. The films employed never reached thicknesses below 18 nm because of the nitrocellulose lift-off technique used \cite{Lassailly1994}.

\begin{figure*}[htbp]
\centering
\includegraphics[width=0.9\textwidth]{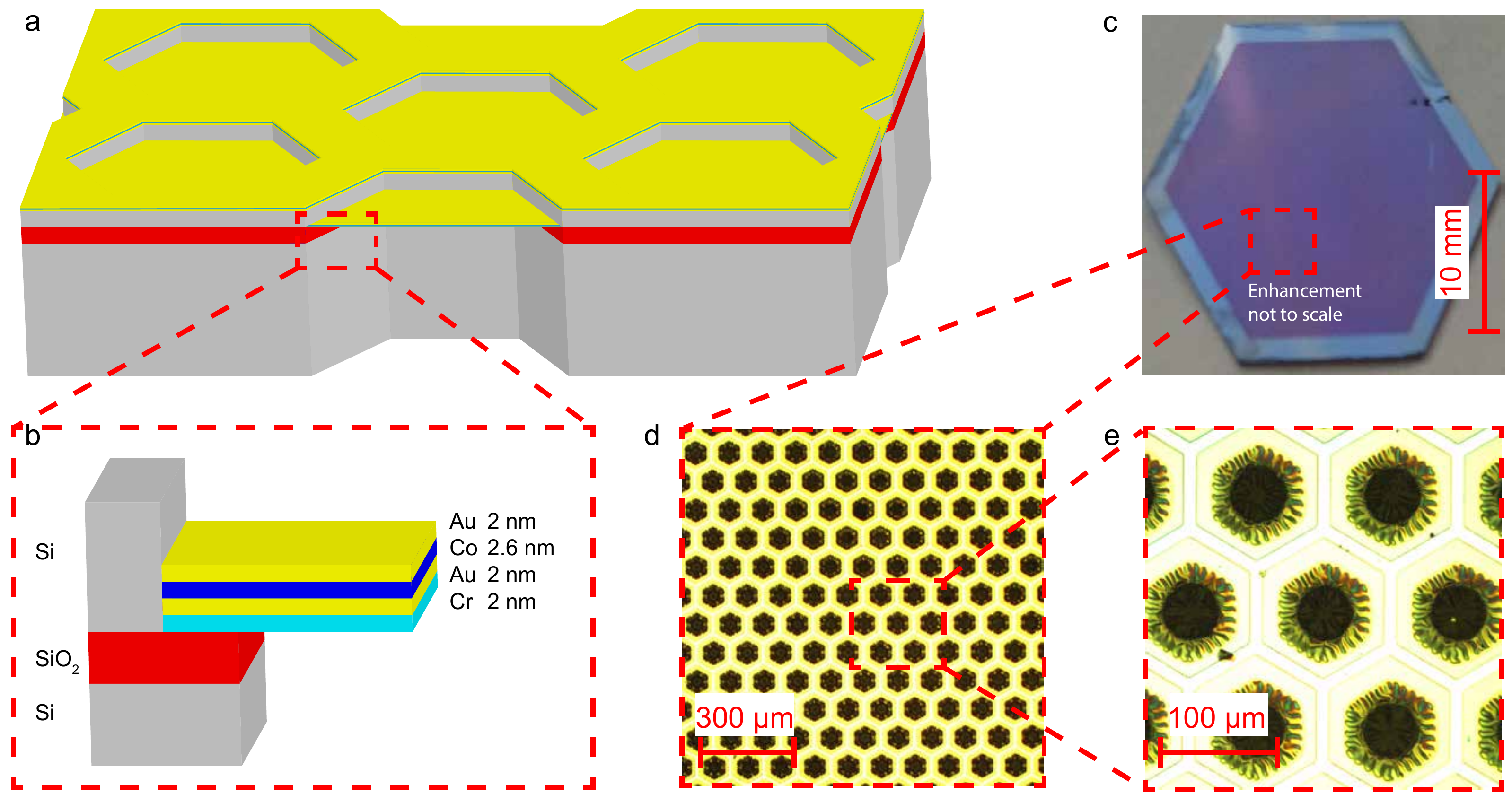}
\caption{\footnotesize Schematic profiles and optical images of the finished filter. a) A schematic profile of the design shows the device supportstructure as the grey (Si) and red (SiO$_2$) suspending the metal membranes. b) A close-up of the metal membrane and layer thicknesses. c) A photograph of the finished chip. d) An image viewed through 5X magnification shows the extensive spread and high yield of the finished membranes. The dark areas are the active, transmitting parts of the device. e) 20X magnification where the details of single membranes can be seen. The dark center of each membrane comprises only free-standing metals. Just off-center a more colorful region due to remaining SiO$_2$ due to quicker etching in the center is visible.}
\label{fig:2}
\end{figure*}

The approach presented in this work, utilizes the rapid development of nanoelectromechanical and microelectromechanical system (NEMS/MEMS) processing technology during the latest decades. The employed Si processing techniques are very versatile, precise and enable elaborate patterns over large areas. By using several sacrificial layers, with high consecutive etching selectivity, it is possible to achieve sub-10 nm free-standing membranes with high uniformity(Fig. \ref{fig:2}). Each such membrane is 80 $\upmu$m wide and a series of them is patterned over an area of several cm$^2$ in the shape of hexagonal chips. The chips in turn are tiled together in front of a multi channel plate enabling massively parallel electron spin detection.

Each chip has a flat-to-flat distance of 17.5 mm comprising over 15000 free-standing membranes (Fig. \ref{fig:2}c). Fig. \ref{fig:2}a describes the chip's fabrication and its final profile. The device starts out as a silicon-on-insulator wafer 300 $\upmu$m thick, which back side is patterned using standard contact lithography. By inductively charged plasma reactive ion etching (ICP-RIE) the Si bulk is removed down to the buried oxide of the wafer which acts as an etch stopping layer. ICP is a deep-etching technique allowing high aspect ratios i.e. depth versus open area. In order to detect when the process has reached the SiO$_2$, an optical emission end-point detection device is used. In the following step, the top Si is patterned and aligned with the bottom pattern using contact lithography. The top Si is then removed using ICP-RIE once again. What remains at this point are hexagonal, 500 nm thick, free-standing SiO$_2$ membranes surrounded by Si bulk. This acts as a support structure for the rest of the device processing. Using magnetron sputtering, polycrystalline metal layers of Cr (2 nm), Au (2 nm), Co (2.6 nm) and Au (2 nm) (in that order) are deposited onto the free-standing SiO$_2$. The different layer thicknesses are controlled by measuring individual deposition rates using X-ray reflectometry. The key layer is Co which is used as the active spin-filtering component. Its homogeniety and in-plane magnetization was controlled using scanning tunneling x-ray microscopy (see supplementary information). Au acts as a capping layer to prohibit oxidation of the Co, as well as mechanical support, while Cr is used as a wetting layer for the Au against the SiO$_2$ surface. Another essential reason for the choice of Cr is its high resistance to standard fluorocarbon chemistry used when removing SiO$_2$ \cite{Williams2003}, enabling very precise processing possibilities. Fig. \ref{fig:2}a and b depicts a schematic profile of the finished device where the only remaining parts are the Au/Co/Au/Cr stack and the Si support structure. In the final device, the Cr is thinner than the initial 2 nm since it acts as an etch-stop layer and is partly consumed. The membranes are finally magnetized using permanent magnets thus enabling the spin selectivity.
Fig. \ref{fig:2}c displays the complete chip and fig. \ref{fig:2}d the extensive spread and homogeneity of the finished membranes as viewed through an optical microscope. The progress of the etching is controlled by observing the gradual consumption of SiO$_2$ through the microscope. This is seen as a surface structural and color change of the membrane, originating from its center (Fig. \ref{fig:2}e). The etching speed will be higher at the center of the membrane thus making the selectivity between Cr and SiO$_2$ extra important to achieve a maximum free-standing area of each membrane as well as an even thickness of the membranes. Since the transmission decreases exponentially, device homogeneity is crucial for an even FOM.

To investigate the transmission, the chips were placed in ultra-high vacuum (UHV) between a multichannel plate detector and an electron source. The energy-dependent transmission is presented in Fig. \ref{fig:1}c together with the "universal curve" for the inelastic mean free path (IMFP) of electrons in solids. The general agreement between the measured transmission and the IMFP is expected but it must be noted that in the measurement, secondary electrons generated in the filter are also detected. This is by choice since secondary electrons contain information on the spin of the primary electron as well \cite{Oberli1998}. As a result, the measured transmission increases faster than the IMFP as the amount of generated secondary electrons increases on the high energy side. As the IMFP becomes comparable to the filter thickness, the amount of secondaries decreases and the measured transmission levels off.

\begin{figure}[htbp]
\centering
\includegraphics[width=0.45\textwidth]{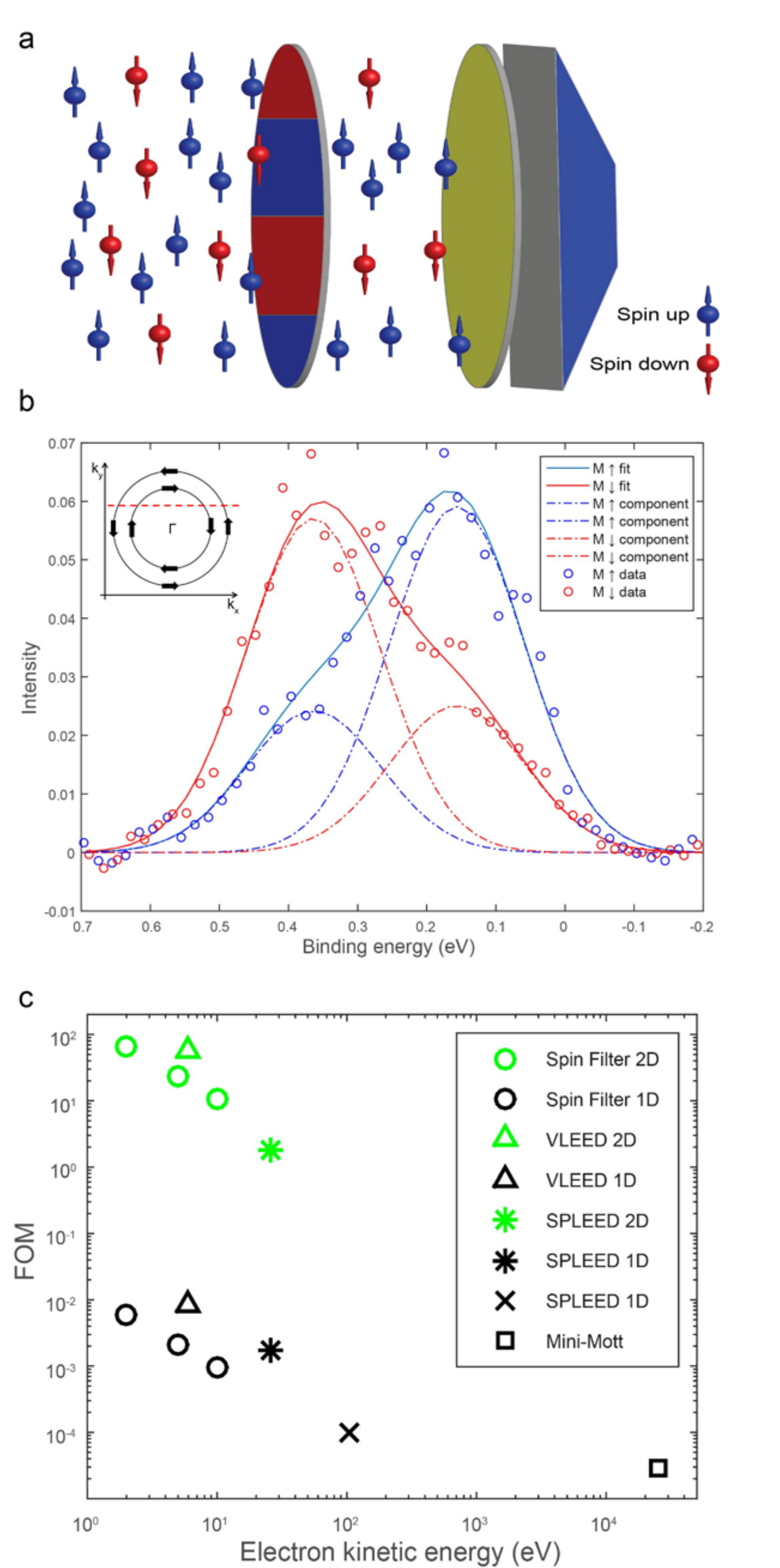}
\caption{\footnotesize Illustration of the experimental setup, spin-ARPES measurement results and FOM comparisons. a) A schematic of the setup used to measure electron transmission and Sherman function. The spin filter is magnetized as stripes, anti-parallel and in-plane. The electron polarization versus the direction of magnetization allows for an asymmetry to be measured (Eq. (\ref{eq:Assym})) which is seen as a difference in contrast on the detector. b) Curve-fitted EDCs from the spin-ARPES measurement, taken at $E_\mathrm{K}=2$ eV. The Rashba split from the surface is used as a polarized electron source and clearly visible as the difference between the red and blue lines. The Sherman function is extracted by curve fitting of the normalized data from the measurement. The top left inset is a schematic image of the surface state. The dashed red line describes where the measurement is taken. c) Comparative chart of $FOM$ and $FOM_\mathrm{2D}$ versus energy for the present spin filter (circles) compared to a few other readily available technologies \cite{Burnett1994,Yu2007,Kolbe2011,Okuda2008,Ji2016,Winkelmann2008}.}
\label{fig:3}
\end{figure}
In order to determine the FOM, it is also necessary to acquire the Sherman function. This is done by performing a spin dependent measurement employing a source of spin polarized electrons. To achieve this, we have performed a spin and angle resolved photoemission spectroscopy (spin-ARPES) measurement on a Au(111) crystal where the reconstructed surface has a Rashba spin split Shockley state \cite{LaShell1996}. This was done at at the
MAX III synchrotron of the MAX-laboratory in Lund, Sweden using a UHV system including a Scienta Omricon R3000 hemispherical electron analyzer installed with a horizontal slit geometry. The light source was p-polarized at $E_{h\upnu}=22$ eV. Since $S$ also depends on the electron energy above the Fermi energy \cite{Weber1999,Oberli1998}, measurements were taken at several kinetic energies, $E_\mathrm{K}=2$ eV, 5 eV and 10 eV as measured at the spin filter. The kinetic energy of the electrons passing through the filter was set by choosing the corresponding pass energy of the analyzer. $E_\mathrm{K}=2$ eV and 10 eV measurements were recorded with the Au(111) sample at $p<5*10^{-9}$ mbar and room temperature while the $E_\mathrm{K}=5$ eV measurement was recorded at $p<5*10^{-10}$ mbar and $T=170$ K. The photo emitted electrons from the spin split parabola of the surface state act as the spin polarized source. The electrons are transmitted through the membranes close to the detector. Magnetizing the membranes result in a change of attenuation length for the electrons depending on the electrons' spin as compared to the direction of magnetization; electrons with spin parallel to the magnetization will have a greater probability of transmission as compared to ones with anti-parallel spin \cite{SiegmannBook}. By magnetizing certain regions of the device (Fig. \ref{fig:3}a) anti-parallel to each other and comparing the transmission of the different regions at a given energy, we obtain the asymmetry by
\begin{align}
A=(I_\uparrow-I_\downarrow)/(I_\uparrow+I_\downarrow)
\label{eq:Assym}
\end{align}
where $A$ is the asymmetry, $I_\uparrow$ is the intensity of majority electrons, and $I_\downarrow$ is the intensity of minority electrons.

The Sherman functions for the different energies were calculated by curve-fitting the energy distribution curves (EDCs) measured at two different regions of the detector with anti-parallel magnetization (see supplementary information). The normalized data and curve-fitting for $E_\mathrm{K}=2$ eV can be seen in fig. \ref{fig:3}b. The spin split of the Rashba state is clearly visible as the difference between the red and blue lines. For a spin integrated measurement, both lines would overlap as the sum of both spin split measurements. At $E_\mathrm{K}=2$ eV, a Sherman function $S\approx 0.41$ was obtained resulting in our highest $FOM=6.0\times10^{-3}$. For comparative reasons, $FOM_\mathrm{2D}$ (previously defined by \cite{Kolbe2011}), is also calculated. This is the product of the number of resolvable channels in energy and angle, $N$, of the spin detector multiplied by the $FOM$ for a single channel, $FOM_\mathrm{2D}=N\times FOM$. The number of channels is determined by dividing the detector area by the experimental resolution. The latter is taken as the the Gaussian broadening obtained by fitting a momentum distribution curve (MDC) with two Lorentzians convoluted by a Gaussian (see supplementary information). The resulting number of channels is  $N=11106$ and the corresponding $FOM_\mathrm{2D}=67.2$ for $E_\mathrm{K}=2$ eV. To the authors' knowledge this is the highest reported to date. It should also be noted that the Sherman functions quoted here are only lower bounds due to the assumption of a completely polarized source. In practice, the polarization is $<1$ due to non-perfect alignment, (Fig. \ref{fig:2}b inset)and surface contamination\cite{Reinert2003,Forster2006,Forster2007,Bentmann2013}. It is also an open question whether the surface state is fully polarized or not \cite{Henk2003}. Typical values in the literature vary from $P\approx 0.4$ to $P\approx 0.8$ \cite{Hoesch2004,Tusche2015}.

Fig. \ref{fig:3}c displays measured $FOM$ and $FOM_\mathrm{2D}$ for electron kinetic energies $E_\mathrm{K}=2$ eV, 5 eV and 10 eV as compared to a few other representative electron spin detectors. As expected, the FOM for the present spin filter increases exponentially, with decreasing electron energy. Comparing the present results with existing technology, we note an increase of the FOM over two orders of magnitude as compared to the widely used Mini-Mott system \cite{Burnett1994} and at par with the far more recently developed very low energy electron diffraction (VLEED) systems \cite{Ji2016,Okuda2008,Winkelmann2008}. When looking further to the $FOM_\mathrm{2D}$, the present results even surpass the 2-dimensional VLEED systems. In addition to this, the VLEED systems are far more complex and require frequent cleaning and regeneration in contrast to the present spin filters that do not require any in situ preparation. This in turn opens up for implementations in fields other than spin-ARPES such as spin-polarized scanning electron microscopy and spin-polarized low energy electron microscopy.

To summarize, we have presented a transmission-based, nano-structured electron spin-filtering device compatible with different spectroscopy and microscopy techniques. The transmission based filter is achieved through NEMS processing techniques and the use of multiple sacrificial etch stops which allowed us to obtain sub-10 nm free-standing metallic membranes. These spin-filters, have a high $FOM$ and permit two-dimensional detection capabilities with the highest $FOM_\mathrm{2D}$ reported to date. The lack of high voltage and no requirements on in-situ preparation makes the present spin filter a compelling alternative to existing technologies. We envision that our results and the presented technology will greatly improve the resolution of spin-resolved electron spectroscopy and microscopy, and also lead to a more widespread use of these techniques.

\begin{acknowledgements}
The authors would like to thank The Swedish Research Council (VR), Swedish Governmental Agency for Innovation Systems (VINNOVA) and the Knut and Alice Wallenberg Foundation for the financial support that enabled this project.
\end{acknowledgements}

\bibliography{Ref}

\newpage

\onecolumngrid

\section*{Supplementary Information}

\setcounter{equation}{0}
\setcounter{figure}{0}
\setcounter{table}{0}
\setcounter{page}{1}

\section{SI.1 - Scanning Tunnelling X-Ray Microscopy}
In order to fully characterize the morphologic and magnetic properties of our spin filter, we used x-ray microscopy and spectroscopy techniques available at synchrotron light sources.

\begin{figure}[b]
\centering
\includegraphics[width=\textwidth]{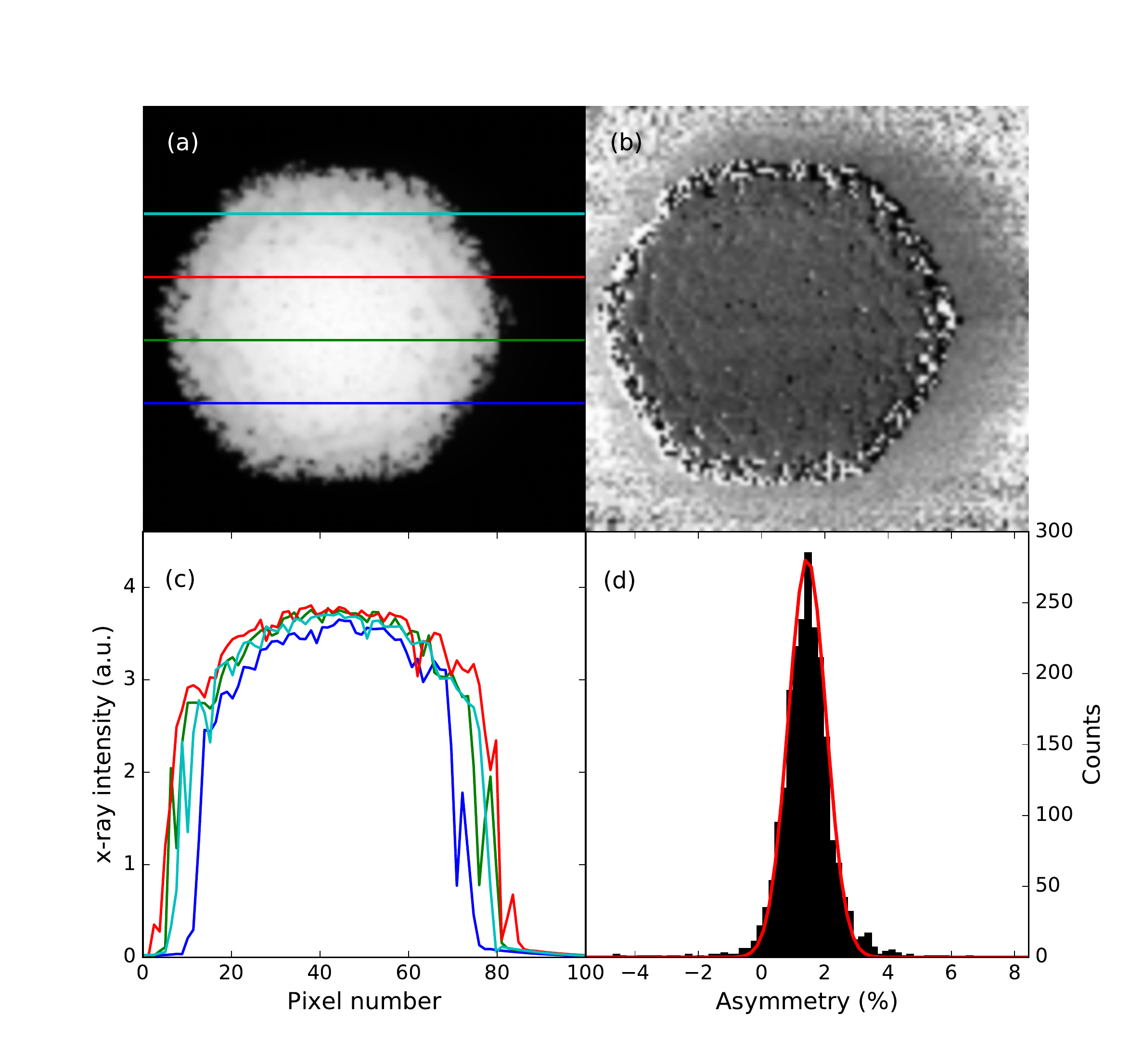}
\caption{(a) Topographic and (b) magnetic contrast obtained from the sum and, respectively, the asymmetry  of STXM images recorded using x-ray radiation with opposite circular polarization. Both images cover an area of 100 $\upmu$m $\times$ 100 $\upmu$m. (c) Line scans across the membrane shown in (a). (d) Histogram showing the distribution of the asymmetry signal in (b) across the membrane.}
\label{fig:S1}
\end{figure}

Given that the spin filter comprises thin free-standing membranes, scanning transmission x-ray microscopy (STXM) is a natural choice of technique. Using the STXM beamline 13-1 at Stanford Synchrotron Radiation Laboratory (SSRL), which is located at the exit of an elliptically polarizing undulator, we could record images of the structures using x-ray light of opposite helicity. Co thin films with a thickness above 12 \AA are spontaneously magnetized in plane.[S1] In order to rotate the magnetization out of plane for the STXM measurements, the spin-filter was subjected to a magnetic field of 0.6 T applied perpendicular to the plane of the membranes.

Summing the two images, we could retrieve the topographic contrast of the spin filter plotted in Fig.~\ref{fig:S1}(a). By taking their difference divided by the sum, the asymmetry of the magnetic contrast becomes visible, as shown in Fig.~\ref{fig:S1}(b). In Fig.~\ref{fig:S1}(c) we plot several line scans taken across the topographic image. Such line scans show a remarkable uniformity of the thickness of the membrane across its surface. Finally, in Fig.~\ref{fig:S1}(d) we plot a histogram of the variation of the asymmetry calculated as $(I^+-I^-)/(I^++I^-)$, where $I^{+(-)}$ is the transmitted intensity of x-rays with positive (negative) helicity. Even in this case, the signal shows good uniformity, and we attribute part of the experimental uncertainty to the drift of the microscope while measuring such large areas.

\begin{figure}[h]
\centering
\includegraphics[width=0.5\textwidth]{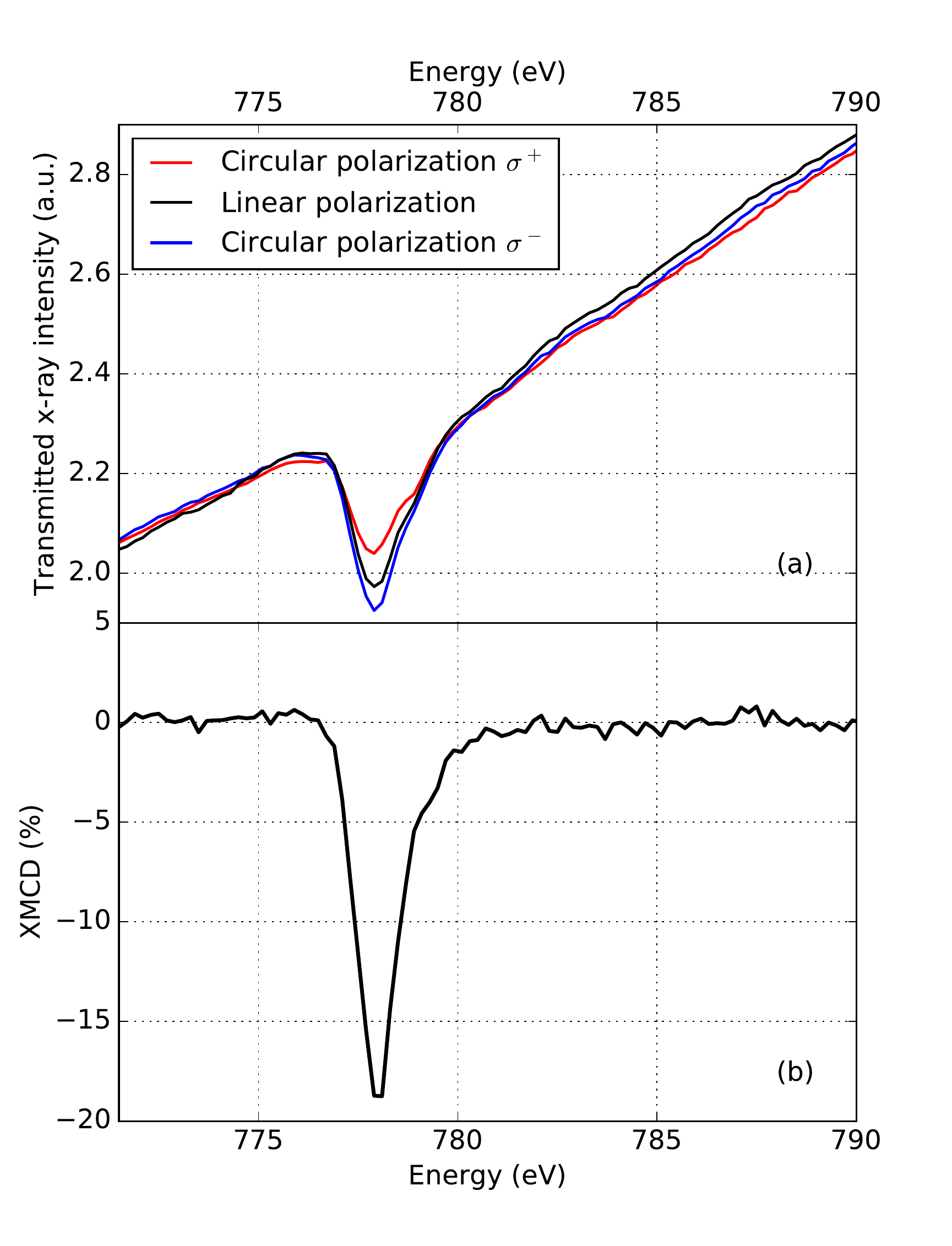}
\caption{(a) Typical x-ray absorption spectra and (b) XMCD spectrum for the membranes imaged with the STXM.}
\label{fig:S2}
\end{figure}

In order to perform quantitative estimates of both thickness and magnetization of the magnetic layer in the membranes, we performed x-ray absorption spectroscopy measurements across the Co $L_3$ absorption edge while the x-ray beam was focused and transmitted through the free-standing membranes. These measurements are shown in Fig.~\ref{fig:S2}(a) for different x-ray polarizations. From the absorption of the linearly polarized light we can retrieve the thickness of the Co layer alone, which is responsible for the absorption change across the edge. Recalling that the transmitted x-ray intensity can be written as
$I=I_0\,e^{-\mu_xt}$, where $I_0$ is the incident x-ray intensity, $\mu_x$ is the absorption coefficient, this allows for an estimation of the thickness $t$ of the material, i.e. $t=-\log{(I/I_0)}/\mu_x$. For Co, $\mu_x\approx5.8\times10^{-2}$ nm$^{-1}$ [S2] and using the measured intensity dip $I/I_0\approx86$\% , this leads to $t_{\rm Co}\approx2.6$ nm.

In order to quantify the magnetization of the sample, one has to calculate the x-ray magnetic circular dichroism (XMCD). This is defined as XMCD $ = {(\mu^+-\mu^-)}/{(\mu^++\mu^-)}$, where $\mu^+$ and $\mu^-$ are the absorption coefficients for positive and, respectively, negative helicity of the x-ray photons. It is convenient to rewrite the expression above in terms of x-ray intensities $I^+$ and $I^-$, i.e. XMCD $=\dfrac{\log I^+/I^-}{\log I^+I^-/I_0^2}$. This quantity is plotted in Fig.~\ref{fig:S2}(b), and shows a value of about 18.5\% at the Co $L_3$ edge. In order to retrieve the magnetic moment of the film, one has to compare this value with the maximum XMCD measurable in a Co film. This can be computed as half of the ratio between the density of Bohr magnetons and the density of holes in the material. Co has 1.6 $\mu_B$/atom and 2.5 holes/atom, leading to a maximum XMCD of 32\% for a fully saturated sample. This indicates that the measured XMCD corresponds to 58\% of the total magnetic moment, consistent with a fully in plane magnetic Co layer whose magnetization is tilted 35 degrees out of plane. Assuming a saturation magnetization $M_s=1.6$ T for bulk Co and solving the magnetostatic boundary conditions, the tilt angle of the magnetization is expected to be about 25 degrees. The additional experimental tilt ($\sim10$ degrees) is likely due to surface anisotropy (expected in such thin Co layers [S1]) which helps reducing the saturation magnetic field in the out-of-plane direction. Use of the Stanford Synchrotron Radiation Lightsource, SLAC National Accelerator Laboratory, is supported by the U.S. Department of Energy, Office of Science, Office of Basic Energy Sciences under Contract No. DE-AC02-76SF00515

\section{SI.2 - Data fitting EDC for Sherman function}
\begin{figure}[H]
\centering
\includegraphics[width=0.5\textwidth]{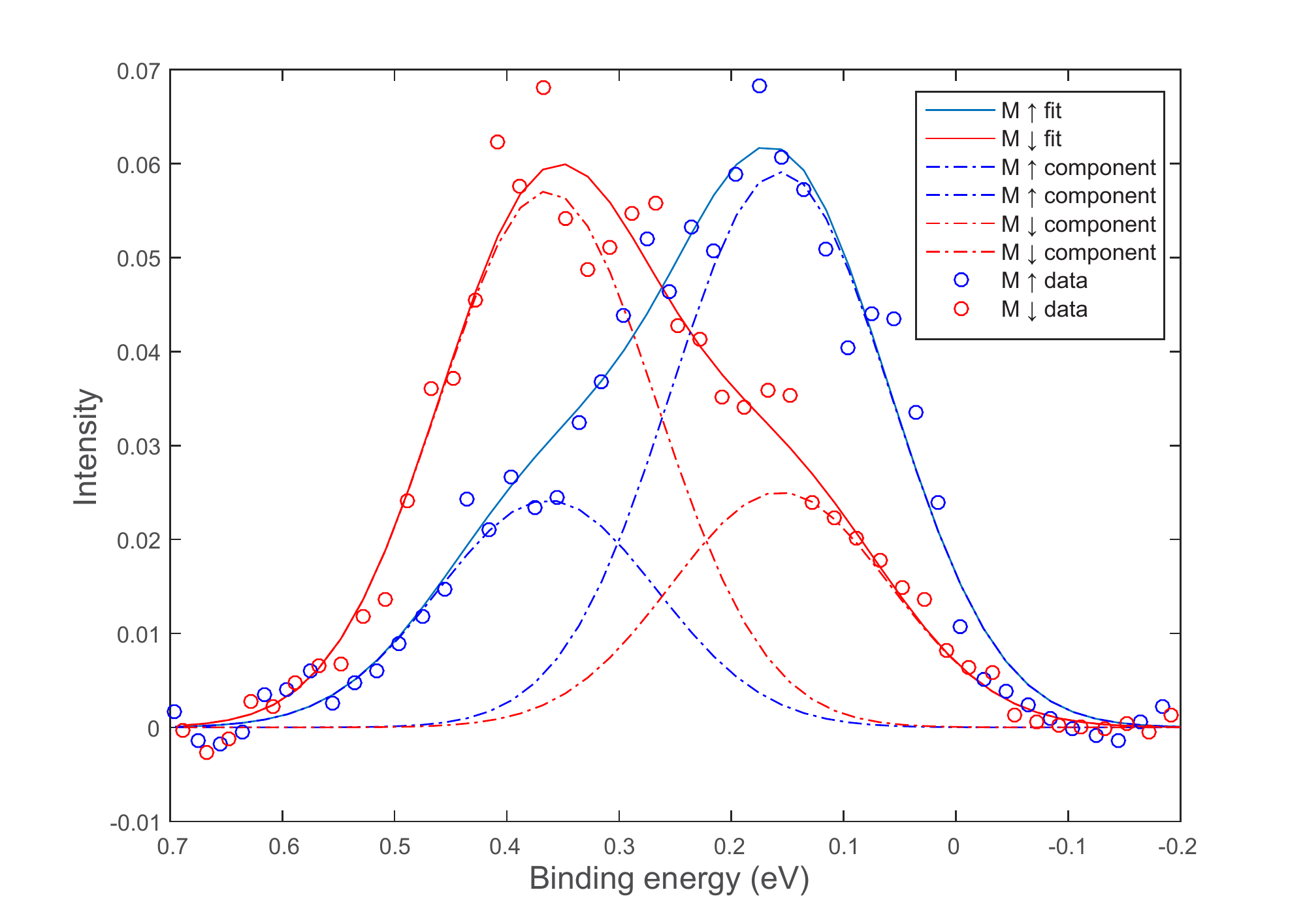}
\caption{Normalized data fitted for $E_\mathrm{K}=2$ eV.}
\label{fig:2eV}
\end{figure}

\begin{figure}[H]
\centering
\includegraphics[width=0.5\textwidth]{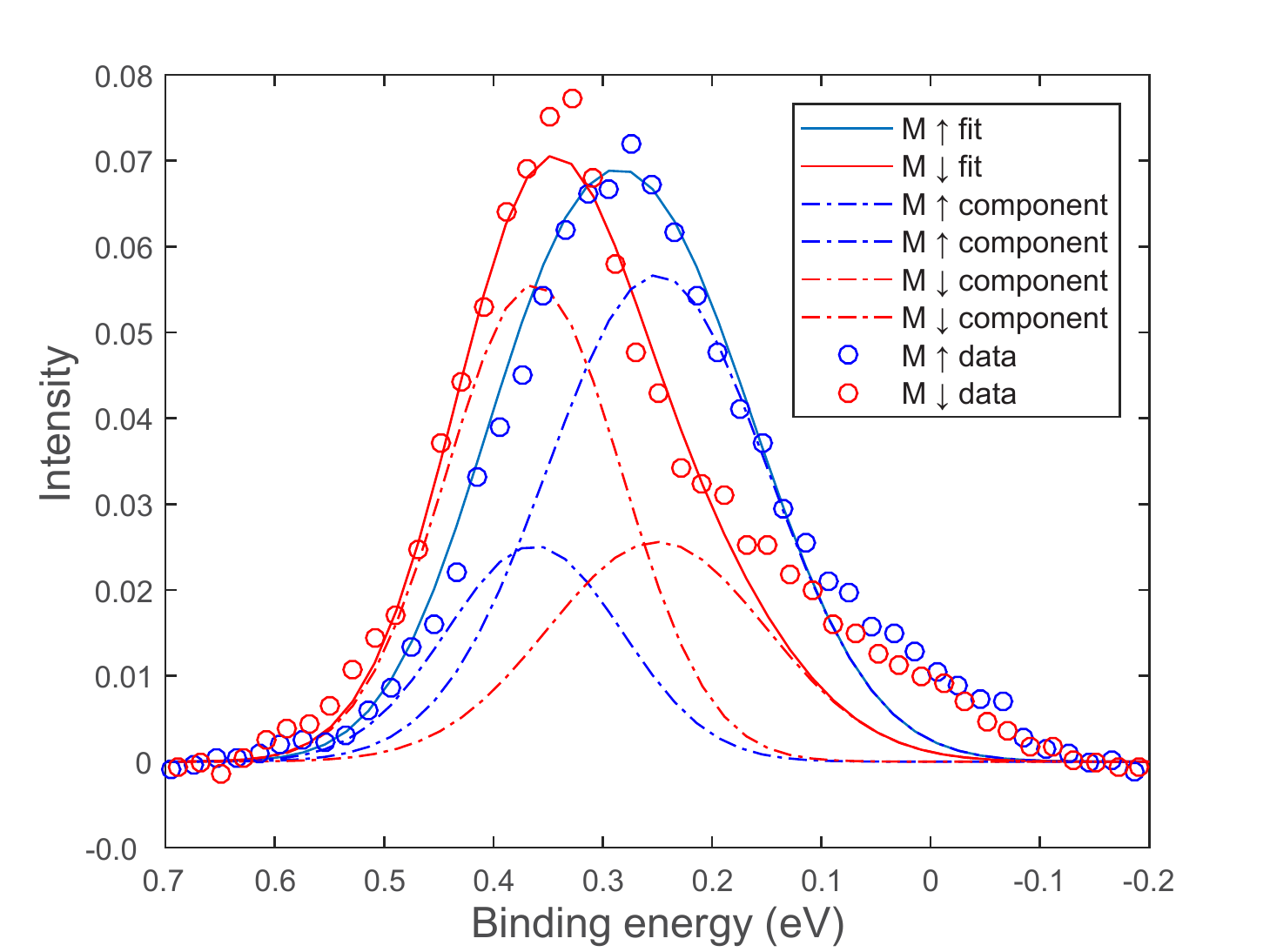}
\caption{Normalized data fitted for $E_\mathrm{K}=5$ eV.}
\label{fig:5eV}
\end{figure}

\begin{figure}[H]
\centering
\includegraphics[width=0.5\textwidth]{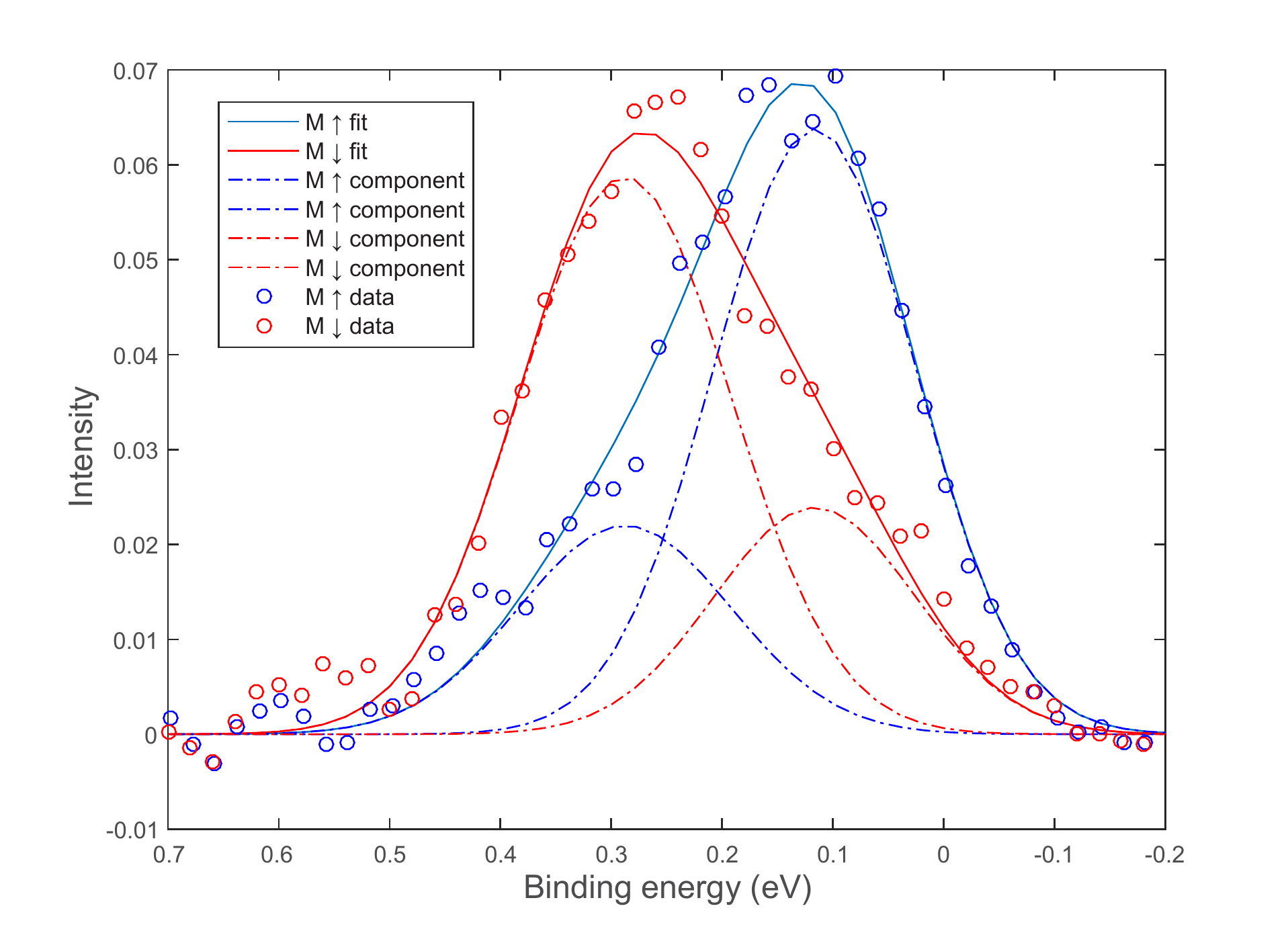}
\caption{Normalized data fitted for $E_\mathrm{K}=10$ eV.}
\label{fig:10eV}
\end{figure}

Data acquired from the angle resolved photo emission spectroscopy (ARPES) measurements of a Au(111) crystal surface is used to estimate the Sherman function of our spin filter for kinetic energies $E_\mathrm{K}=$ (2, 5 and 10) eV. Energy distribution curves (EDCs) from two regions of the detector with anti-parallel magnetization are fitted with the sum of two Gaussian functions, $M_\uparrow=I_1Gauss(x_1,\sigma_1)(1+SP_1)+I_2Gauss(x_2,\sigma_2)(1+SP_2)$ and $M_\downarrow=I_1Gauss(x_1,\sigma_1)(1-SP_1)+I_2Gauss(x_2,\sigma_2)(1-SP_2)$ using a least squares criterion. The parameters used are peak intensities (I) and positions (x), width ($\sigma$), peak polarization (P) and Sherman function (S). When fitting the two differently magnetized regions, only the intensities are allowed to vary between the two, all other parameters are kept fixed. The polarizations are assumed to be 1 and -1 and the Sherman function is therefore defined as $S=(I_1-I_2)/(I_1+I_2)$. Prior to fitting, the data is normalized using integral background subtraction. The individual fits and corresponding parameters are presented in Fig. \ref{fig:2eV}-\ref{fig:10eV} and Table \ref{tab:table1}. The difference in energy split when comparing results between the different electron kinetic energies can be explained by surface adsorbates.

\begin{table}[hbt]
\begin{ruledtabular}
\begin{tabular}{cccccccccc}
$E_\mathrm{K}$ (eV) & $I_1$ & $x_1$ (eV) & $\sigma_1$ (eV) & $I_2$ & $x_2$ (eV) & $\sigma_2$ (eV) & $P_1$ & $P_2$ & $S$ \\
2 & 0.042 & 0.16 & 0.097 & 0.041 & 0.36 & 0.097 & 1 & -1 & 0.41\\
5 & 0.041 & 0.25 & 0.10 & 0.040 & 0.36 & 0.080 & 1 & -1 & 0.37\\
10 & 0.044 & 0.22 & 0.091 & 0.040 & 0.39 & 0.096 & 1 & -1 & 0.45
\end{tabular}
\end{ruledtabular}
\caption{Parameters extracted from data fitting of EDCs at different $E_\mathrm{K}$.}
\label{tab:table1}
\end{table}

\section{SI.3 - Data fitting MDC for resolution }
\begin{figure}[H]
\centering
\includegraphics[width=0.5\textwidth]{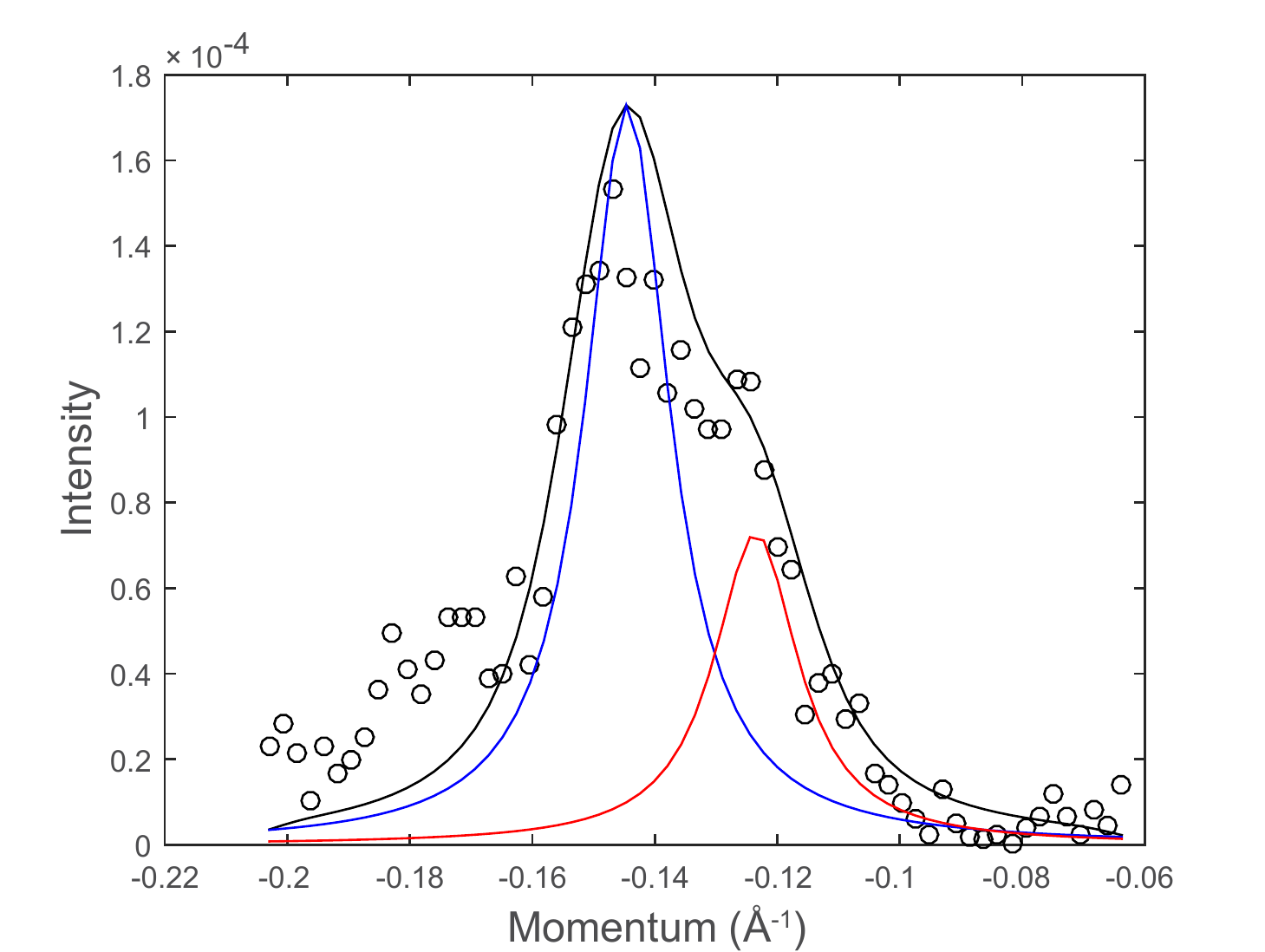}
\caption{MDC data fitted (black) for $E_\mathrm{K}=5$ eV and Lorentzians (blue and red) for fitting.}
\label{fig:MDC}
\end{figure}

To calculate the $FOM_\mathrm{2D}$ the number of single resolvable channels needs to be estimated. This is done by looking at the momentum distribution curves (MDCs) close to the Fermi level from the data collected at electron energies of $E_\mathrm{K}=5$ eV and fitting them to the sum of two Lorentzians, representing the spin split parabolas of the Au(111) surface, convoluted with a Gaussian broadening. The Lorentzian intensities are normalized and weighed against the Sherman function according to $I_1 = 1$ and $I_2 = I_1\dfrac{(1-S)}{(1+S)}$. The spin split data is taken from [S3] and the Sherman function (S) comes from our own experimental data. Using a least squares fit, the broadening ($\sigma$) and intensity (I) of the Gaussian is fitted to the data. The resulting $F.W.H.M. = 2\sqrt{2ln(2)}\sigma$ is used as the resolution and thereby the number of single channels on the detector, $N = \dfrac{ detector width}{F.W.H.M.} \times \dfrac{detector height}{F.W.H.M.}$ where the broadening is assumed as geometrically symmetric at the spin filter. The calculations are done for $E_\mathrm{K} =$ 5 eV and the MDC data and fit are presented in Fig. \ref{fig:MDC}. The fitted Gaussian parameters obtained are $\sigma_\mathrm{G} =$ 0.0052 \text{\AA}$^{-1}$ and I$_\mathrm{G} =$ 0.00017.

\par\noindent\rule{\textwidth}{0.4pt}
\section*{}
[S1] P. Bruno and J.-P. Renard, J.P. Appl. Phys. A (1989) 49: 499. 

[S2] J. St\"{o}hr and H.C. Siegmann, Magnetism: From Fundamentals to Nanoscale Dynamics. vol. 152. (Springer, 2007)

[S3] M.H. Berntsen, O. G\"{o}tberg and O. Tjernberg, Review of Scientific Instruments 82, 095113 (2011)
\end{document}